# The GōMartini approach: Revisiting the concept of contact maps and the modelling of protein complexes

*Luis F. Cofas-Vargas[1], Rodrigo A. Moreira[2], Simón Poblete[3,4], Mateusz Chwastyk[5], Adolfo B. Poma*[1]*

[1]Biosystems and Soft Matter Division, Institute of Fundamental Technological Research, Polish Academy of Sciences, ul. Pawińskiego 5B, 02-106 Warsaw, Poland.
[2]BCAM, Basque Center for Applied Mathematics, Mazarredo 14, 48009 Bilbao, Bizkaia, Spain
[3]Centro Científico y Tecnológico de Excelencia Ciencia & Vida, Fundación Ciencia & Vida, Avenida del Valle Norte 725, 8580702 Santiago, Chile
[4]Facultad de Ingeniería, Arquitectura y Diseño, Universidad San Sebastián, Bellavista 7, 8420524 Santiago, Chile
[5]Institute of Physics, Polish Academy of Sciences, al. Lotników 32/46, 02-668 Warsaw, Poland

*Author for correspondence:
Adolfo B. Poma (E-mail: apoma@ippt.pan.pl)

## Abstract

We present a review of a series of contact maps for the determination of native interactions in proteins and nucleic acids based on a distance-threshold. Such contact maps are mostly based on physical and chemical construction, and yet they are sensitive to some parameters (e.g. distances or atomic radii) and can neglect some key interactions. Furthermore, we also comment on a new class of contact maps that only requires geometric arguments. The contact map is a necessary ingredient to build a robust GōMartini model for proteins and their complexes in the Martini 3 force field. We present the extension of a popular structure-based Gō-like approach for the study of protein-sugar complexes, and also limitations of this approach are discussed. The GōMartini approach was first introduced by Poma *et al. J. Chem. Theory Comput.* 2017, **13(3), 1366-1374** in Martini 2 force field and recently, it has gained the status of gold-standard for protein simulation undergoing conformational changes in Martini 3 force field. We discuss several studies that have provided support to this approach in the context of the biophysical community.

# 1. Introduction

Structural biology has made significant strides in recent years, fueled by advancements in experimental techniques like nuclear magnetic resonance (NMR), X-ray crystallography, and cryo-electron microscopy (cryo-EM). These techniques provide detailed insights into the three-dimensional structures of biomolecules, shedding light on their functional mechanisms. However, static structural data alone fails to capture the dynamic aspects of molecular biology. To bridge the gap between static structural data and dynamic experimental data, robust and versatile computational models capable of accurately describing the dynamics of biomolecular complexes are essential. The GōMartini approach[1] for proteins offers versatility by combining the latest Martini 3 force field[2] for proteins and other biomolecules (e.g. lipids, carbohydrates, nucleic acids, etc) and its cost-effective edge renders this approach ideal for large-scale applications in cellular environments[3].  Structure-based (SB) model offers a promising approach, utilising coarse-grained (CG) representations to capture the essence of a biomolecule structure and dynamics.

The typical time scales of biological processes involving e.g. unfolding of proteins, protein recognition, among other events are in the range of $10^{-6}$–$10^{-3}$ s, and thus they are orders of magnitude slower than typical molecular motion (i.e. $10^{-15}$–$10^{-12}$ s) simulated in all-atom (AA) molecular dynamics (MD). The length scales of conformational rearrangements are also much smaller in AA-MD simulation than they would be for studying processes involving large structural changes in biological systems. In this regard, the SB model and CG approaches of biomolecular systems are ideal tools to overcome such limitations. The replacement of the position of each amino acid by its $C^{\alpha}$ atom is a common choice. In this approach, several degrees of freedom of the system are removed, which enables reaching the experimental time and length scales, while maintaining a molecular-level model of the systems under consideration. In particular, CG approaches are used to infer Young modulus and confront it with atomic force microscopy (AFM) experiments. Importantly, the mechanism of deformation that gives rise to the linear force-displacement response can be characterised in the CG simulation. Several CG models are not sensitive to pH, ionic strength, and they also do not consider the electrostatic interactions, post-translational covalent modifications of amino acids, etc. Those factors have been demonstrated to be important in, for example, recognition of cell receptors by pathogens and control of the assembly of protein complexes. In addition, standard AA-MD simulation can target system sizes on the scale of ~500 Million atomistic particles in the latest SARS-CoV-2 full virion in aerosol droplet[4], which is only possible in a few high-performance computing clusters around the world. However, analogous systems formulated using CG force fields such as Martini 3[2], SIRAH[5], UNRES[6], are an order of magnitude smaller. In CG-MD simulation, those systems can be studied in a moderate-size computing cluster. Moreover, due to the large time-step used in CG-MD (e.g. MD simulation with Martini 3 employs $\Delta t_{CG}$ = 20 fs in comparison to AA-MD with a $\Delta t_{MD}$ = 2 fs), thus CG simulations are expected to reach longer time scales than AA-MD. At the core of the SB approach in proteins lies the concept of native interactions, also known as "native contacts" (NC), which provides a simple form to understand the important interactions in equilibrium; it represents the close spatial proximity between residues or atoms in the native state. Defining native interactions poses a challenge, as simple cutoff distance-based definitions can lead to two incompatible outcomes: i) the exclusion of relevant contacts beyond 6 Å, and ii) the introduction of nonphysical next-nearest neighbour contacts. To address these limitations, various methods have been developed to define native contacts, including atomic overlap map, shadow map, CSU contact map, and Voronoi maps (to be discussed in the next section). Each method offers unique advantages and limitations, and the optimal choice depends on the specific application. In the past, we combined both the semi-atomistic approach (e.g. Martini 3 force field) and the SB approach and as such we developed an alternative strategy to study conformational changes of proteins and through this review work we plan to show the extension to protein complexes.

Hence, here we discuss first several contact maps that employ distance cut-off, chemical and physical information, secondly we briefly introduce the extension of the popular SBM developed for protein-sugar complex and lastly the more robust model, the so called GōMartini approach that is based on the SB model of proteins developed by Prof. Marek Cieplak (e.g. a $C^\alpha$-based Gō-like model) and others. This simple model turned out to be efficient to capture the long-time behaviour of certain biomolecular systems under mechanical forces and under high temperatures[7–11]. Most importantly, the Martini force field with an almost atomic resolution can use a backmapping protocol to recover an AA representation from the CG representation with an almost atomic resolution.

## 2. Contact maps for determination of interaction and topological aspect in proteins and nucleic acids

*2.1 Contact maps based on distance-threshold and geometric principles*

A simple protein contact map (CM) based on a distance cut-off that allows for the calculation of protein interactions depends essentially on the atomic positions (see Fig. 1). This method considers interacting any pair of atoms in different residues that are within a certain distance of each other. For example, in protein studies, contacts have often been defined based on atomic geometry, by selecting the heavy atoms in a given amino acid residue, then an atomic contact is found, if two heavy atoms associated to distance residues are within a specific cutoff distance (i.e. 4.0-6.5 Å)[12]. Despite its simplicity, the cut-off CM suffers from several issues that render it less accurate and reliable for determination of native contacts[13], especially in the context of SB models that require the accurate determination of the native contacts to examine the emerging protein dynamics from the underlying geometry. One of the problems with the cut-off CM is a high sensitivity to the cutoff distance. This means that even slight adjustments to this parameter can result in substantial changes in the number of identified contacts. This sensitivity can impede the comparison of results across different studies. Additionally, the cut-off CM often identifies contacts between atoms that are not physically in contact with each other, leading to erroneous conclusions about molecular structure and function. Furthermore, it fails to account for occluded contacts or structural elements, overlooking the accessibility of atoms and their embedding within the larger-scale structure, potentially overestimating the number of identified contacts.

In contrast, the shadow CM[14] fixes some of the previous limitations of cut-off CM and offers a more advanced approach for determination of atomic contacts within a protein. It considers the concept of "shadows" cast by other atoms. In this method, two atoms are only considered to be in contact if there are no other atoms blocking the line of sight between them. The process of obtaining contacts involves the following steps: i) calculates the distances between all pairs of atoms in the protein and create a list of pairs within a specified cut-off distance, ii) utilises a spherical screening radius, typically ≤ 0.5 Å, centred at each atom, and for each pair of atoms, exclude contacts if one atom is obscured by the shadow of another atom and intuitively captures only the "visible" atoms from the perspective of a reference atom. This procedure is visually depicted in Fig. 1. The Shadow CM has demonstrated superior accuracy compared to the cut-off CM for several reasons. First, it exhibits less sensitivity to the choice of cut-off distance, as this parameter is primarily used to establish the initial set of potential contacts and the occluded contacts are subsequently removed. Second, the shadow CM aligns better with experimental data by capturing contacts between atoms separated by intervening atoms, such as water molecules. Despite its advantages, the shadow CM does come with certain limitations. Notably, it introduces increased computational complexity and retains some sensitivity to the cutoff distance, albeit to a lesser extent than the cut-off CM. This sensitivity to noise may result in the identification of false positive contacts, particularly in molecules with flexible structures (i.e. loops and coils) or when dealing with noisy experimental data. Furthermore, the shadow CM cannot capture solvent-mediated contacts or indirect contacts in general.

An entirely different strategy for defining protein contacts involves a cut-off-free methodology that relies solely on geometric principles[12,15,16]. The Voronoi tessellation[17] is a technique for partitioning the physical space into convex polyhedrons, called Voronoi cells, with each cell associated with a specific site, typically the C$^α$ atom of each amino acid is chosen in the context of protein structure analysis. The Delaunay triangulation[17] complements Voronoi tessellation by connecting a set of points with a network of triangles, ensuring that no point lies inside the circumcircle of any triangle. In the case of protein structure analysis, Delaunay triangulation is often used in conjunction with Voronoi tessellation to define protein contacts. To define these contacts using Voronoi tessellation and Delaunay triangulation, the following steps are typically followed: i) Construct the Voronoi tessellation of the protein structure. ii) For each pair of adjacent Voronoi cells, determine whether the corresponding Delaunay edge exists. iii) If the Delaunay edge exists, then the two sites are considered to be in contact. There are several advantages while using Voronoi tessellation and Delaunay triangulation in protein structure for contact definition and topological properties studies. This method is computationally efficient, robust to data noise, and capable of capturing both direct and indirect contacts.

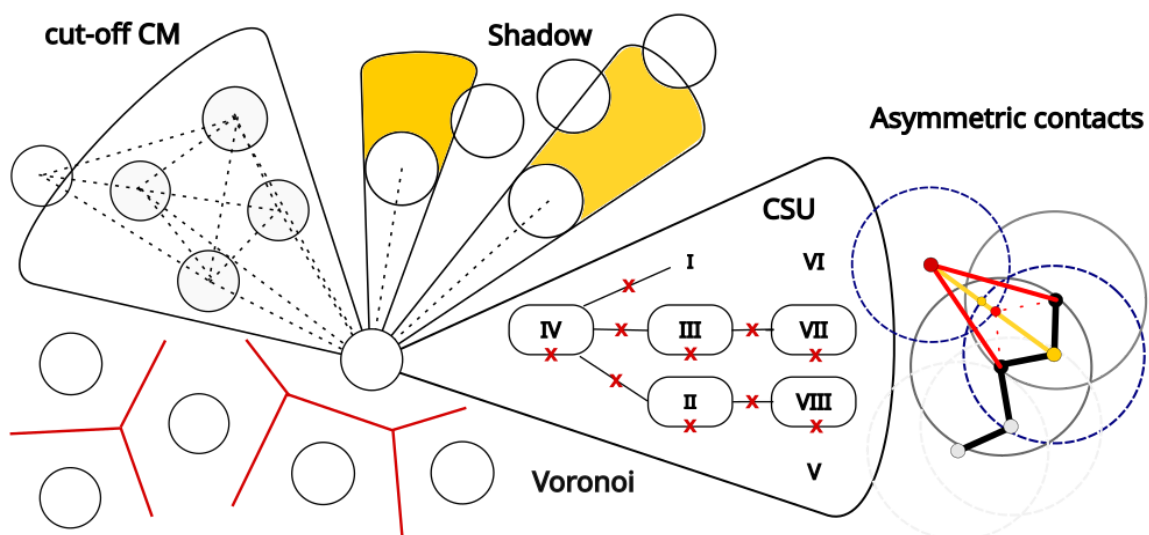

**Figure 1**: Representations of contact maps (CM) in the literature. The cut-off CM used only the distance between centres, while the Shadow methodology improved the former by not discarding centres in the yellow areas. The CM can be improved by including structural categories, the CSU methodology, represented by the roman numbers, and defining illegitimate (marked with red crosses) interaction. However, the CSU CM uses an extended sphere that accounts for solvent effects (dashed circles at right) as a first guess for contacts, which can be asymmetric (as shown by the red and yellow lines at right) due to shadowing effects. Cut-off free pure geometric strategy can use a Voronoi diagram, as shown in the panel below, to create the CM based on the Delaunay triangularization.

*2.2 Contact maps based on chemical and physical information*

The contacts of structural units (CSU) method[18], is a structure-based approach that leverages geometric and chemical information to identify contacts between amino acid residues within a protein. It involves three main steps: i) Identifying pairs of heavy atoms that are in close proximity, typically within a defined distance threshold. ii) Assigning each atom to a specific class based on its chemical properties, such as its element (O, N, C, S) and its connectivity to other atoms. iii) Establishing contacts between residues based on the presence of specific interactions between their individual atoms, including hydrogen bonds, aromatic interactions, and hydrophobic interactions. Any interactions that do not fit into these specific categories, labeled as "non-specific contacts," are excluded from consideration in the CSU method. This exclusion is based on the premise that these non-specific interactions may not convey the structural or functional relevance that the method aims to capture. In essence, the CSU method focuses on recognizing and emphasizing interactions with well-defined chemical characteristics, enhancing the specificity and relevance of the identified contacts within the protein structure. However, the CSU method has certain limitations such as, it only accounts for attractive interactions and neglects repulsive interactions, potentially leading to the inclusion of contacts destabilised by repulsive forces. Additionally, this method may identify contacts between residues that are not physically in contact, as it relies on the presence of at least one specific contact, and also can return asymmetric contacts, as depicted in Fig. 1, due to shadowing effects of neighbouring units. Moreover, it might miss important contacts, particularly in helical structures, due to its focus on specific interactions. Overall, the CSU method provides a valuable approach for identifying contacts in proteins, but it does have limitations stemming from its exclusion of repulsive interactions and potential lack of selectivity.

The repulsive CSU (rCSU) methodology[7] addresses these shortcomings by incorporating repulsive interactions and refining contact identification, offering a more accurate and reliable approach to a new form of contact map generation. The rCSU methodology extends the CSU approach by considering repulsive interactions between charged atoms. It aims to provide a more precise representation of inter-residue contacts by accounting for both attractive and repulsive forces. The rCSU algorithm proceeds in a manner similar to CSU: i) Initially identifying pairs of heavy atoms in close proximity. ii) Subsequently, it classifies atoms based on their chemical properties, akin to CSU. iii) The determination of whether there is a contact between two residues is dependent on the overall balance or net outcome of interactions at the atomic level, calculated as the difference between the number of attractive contacts (e.g. hydrogen bonds, aromatic interactions, ionic bridges and hydrophobic interactions) and the number of repulsive contacts (Coulombic repulsions between charged atoms). iv) If the net contact is positive, a contact between the residues is established. The rCSU methodology offers several advantages over CSU. It provides more accurate contact predictions by considering repulsive interactions, reducing the likelihood of contacts destabilised by repulsive forces. Furthermore, it enhances contact selectivity by evaluating the net contact between residues, lowering the probability of false positives. This methodology also captures a wider range of interactions, including ionic bridges, resulting in a more comprehensive representation of inter-residue contacts. In summary, the rCSU methodology presents a more accurate and reliable approach to CM determination compared to CSU, as it incorporates more chemical information and improves contact selectivity.

The OV+rCSU method[7] combines the strengths of the overlap (OV) method and the rCSU method to identify contact maps in proteins. The OV method identifies contacts based on the overlap of enlarged van der Waals spheres around the heavy atoms, while the rCSU method incorporates repulsive interactions between atoms with charges to refine contact identification. In contrast, the shadow CM method relies on a fixed distance cut-off, independent of atomic size, and removes contacts with intervening atoms. OV+rCSU is superior to the shadow CM method because it considers atomic sizes derived from experimental studies[19] and repulsive interactions, enabling the capture of a broader range of interactions while maintaining selectivity and decrease in false positives.

The CSU method is simpler, only considering attractive interactions and disregarding repulsive interactions. This method can simply lead to false positives, as some contacts may be destabilised by repulsive forces. OV+rCSU addresses this limitation by incorporating repulsive interactions to refine contact identification. While rCSU is an improvement over CSU, it may miss some true contacts due to its focus on net contact between residues. The OV+rCSU method complements rCSU by considering overlaps of van der Waals spheres, potentially capturing additional contacts. OV may identify false positives due to its reliance on overlaps without considering repulsive interactions. rCSU and OV+rCSU address this limitation by incorporating repulsive interactions to refine contact identification.

The Voronoi/Delaunay[15] methodology provides a cut-off-free approach[12] to CM determination, relying on geometric constructs to define contacts based on the proximity and connectivity of residues. It involves partitioning space into polyhedra, known as Voronoi cells, with each cell associated with a residue. The faces of these polyhedra define the closest contacts between residues, offering a geometric foundation for contact definition. This methodology does not require a fixed cut-off distance, eliminating the need for arbitrary cut-off/parameter selection. It delivers a geometric basis for contact definition, ensuring consistency and robustness, capturing both local and global contact patterns, and providing a comprehensive view of the protein structural connectivity. This method can be used to define both residue-residue and atom-atom contacts, offering flexibility in granularity.

The choice between the OV+rCSU and Voronoi/Delaunay methodologies depends on the specific application's requirements for accuracy, efficiency, and robustness. For applications demanding high accuracy and comprehensiveness, such as protein folding simulations or detailed structural analysis, the OV+rCSU methodology may be the preferred recommended choice in SB models. The explicit consideration of atomic sizes and repulsive interactions provides a more detailed and realistic representation of native contacts in proteins. For applications requiring a fast and efficient method for capturing local and global contact patterns, such as network analysis or large-scale structural comparisons, the Voronoi/Delaunay methodology may be the better choice. Its cut-off-free nature and geometric foundation make it computationally efficient and less sensitive to arbitrary cut-off selections. In general, the OV+rCSU methodology is well-suited for applications where a high level of accuracy and detail is crucial, while the Voronoi/Delaunay methodology is well-suited for applications where efficiency and robustness are primary considerations.

Overall, the OV+rCSU methodology offers a more accurate and comprehensive approach to contact map determination compared to rCSU, OV, and Shadow CMs individually. It combines the strengths of the OV and rCSU methods to identify a broader range of interactions while maintaining selectivity and reducing the number of false positives.

A recent method that considers the equilibrium dynamics of a protein, such as the differential/dynamic contact map (dCM)[20] offers an alternative solution. It can identify the most structurally relevant contacts in a protein using AA-MD simulations. This method relies on contact frequency and definition of stability. Frequency measures the number of times a contact was observed between two residues. High contact frequencies indicate more stable contacts. Stability, on the other hand, is determined by considering the chemical characteristics of residues involved in a contact. For instance, hydrophobic interactions are generally more stable than polar-polar and electrostatic interactions. To obtain a more detailed view of the set of protein contacts, the OV+rCSU approach is used with the dCM analysis. The OV+rCSU considers the chemical character of each residue and the respective contacts between a pair of residues, classifying them into categories to count the number of stabilising and destabilising contacts per residue, defining a contact when both residues have a net stabilising character. The dCM and OV+rCSU methodologies together form a robust contact map technique known as differential contact map that has been validated in the study of the dynamics of large protein complexes. For example, the dCM analysis identifies the high-frequency (>0.9) contacts between amino acids in the SARS-CoV-2 trimeric spike protein[20]. It reveals that flexible loops are the source of contact fluctuations, comprising approximately 1772 amino acids based on secondary structural analysis, while helices and strands are roughly represented by 712 and 819 residues, respectively. The entire spike protein has 3363 residues. Indicating that the methodology is feasible even for large protein complexes.

*2.3 Contact maps for intrinsically disordered proteins*

Creating a contact map for intrinsically disordered proteins (IDPs) presents challenges due to their lack of well-defined tertiary structure, which evolves over time. Furthermore, the energetic landscape of these proteins significantly differs from those with stable structures that possess a singular energetic minimum[21], as opposed to the shallow energetic wells between which the protein's conformation fluctuates[22]. This necessitates defining the contact map temporally, updating it at every simulation step. Given the absence of a fixed protein structure, a specialized algorithm is essential for determining this contact map.

One feasible approach is an algorithm based on three criteria: distance between amino acids, orientation of specific residues' side groups, and the potential number of contacts a given residue can establish. The algorithm categorizes contacts into three types: sidechain-sidechain (ss), backbone-backbone (bb), or backbone-sidechain (bs), each utilizing slightly different criteria.

The distance criterion serves as the foundational parameter, governing the onset of a contact. Contacts break when the distance between the centers of $C^{\alpha}$ atoms of particular residues exceeds a defined limit, $f\sigma_{i,j}$, where $\sigma_{i,j}=r_{min} \cdot (0.5)^{1/6}$. $r_{min}$ indicates the position of the energetic potential minimum. The values of $r_{min}$, determined from an analysis of distances between residues in 21,090 non redundant proteins from the CATH database, varied based on the interaction type. The $r_{min}$ values for bb and ss contacts were determined as the mean values from the collected data, resulting in 5.0 Å and 6.8 Å for bb and bs contacts, respectively. However, for $ss$ contacts, $r_{min}$ was individually calculated for each pair of residues. This value ranged from 6.42 Å for the Ala-Ala interaction up to 10.85 Å for the Trp-Trp interaction, and comprehensive details about other pairs of residues capable of forming contacts are presented in ref.[23,24]. As mentioned earlier, contacts fluctuate during the simulation, and a contact is broken when the distance between residues exceeds $f\sigma_{i,j}$, where $f$=1.5. Different values of this factor were also considered and are well described in ref.[23,24]

Another crucial criterion is the orientation of residues. Implementing this criterion is not straightforward as each residue is treated as a spherical bead. Therefore, neighboring residues must be considered for direction implementation, as detailed in ref.[23,24]. Determining the orientation of a backbone hydrogen bond or a sidechain $C^{\beta}$ atom relies on the positions of three consecutive $C^{\alpha}$ atoms. This criterion is essential because we assume that a bb contact can occur if the N-atom on the backbone part of the *i*th residue can establish a hydrogen bond with the O-atom on the backbone part of residue *j*, or vice versa. However, this interaction is permissible only when both atoms are oriented toward each other. The same requirement for residue orientation applies to ss and bs contacts. Detailed descriptions of the mathematical formulas that enable the implementation of these requirements are provided in ref.[23,24].

The final criterion involves the residue types that are essential for defining contacts. They are categorized into six classes: (1) Gly, (2) Pro, (3) hydrophobic, (4) polar, (5) negatively charged, and (6) positively charged. The solvent being implicit in the program restricts the simulation of interactions between polar residues and water molecules. Employing the one-bead-per-residue model leads to a less dense protein representation. To compensate for this, we restrict each amino acid's capacity to form a limited number of contacts. The formula $z_s=n_b+min(s,n_H+n_P)$ determines this number, where '$n_b$' signifies the allowable count of backbone contacts, '$s$' represents the maximum quantity of sidechain contacts, '$n_H$' denotes the upper limit for contacts with hydrophobic residues, and '$n_P$' signifies the limit for contacts with polar side chains. Detailed values for these parameters can be found in ref.[23].

It's crucial to note that the contact map, irrespective of its method of creation, can be utilized with any potential energy function, whether it's a spherical potential like Lennard Jones or one that integrates directional criteria. The initial step always involves validating the distance criterion, followed by evaluating the ability to create specific contacts. The above-mentioned methods for defining contacts is necessary not only for the description of the dynamics of single protein chains but also for research on the aggregation of IDPs or even the creation of protein droplets.

*2.4 Contact maps for nucleic acids: RNA structures*

SB models based on Gō-like approach have also been used to study RNA molecules in CG descriptions. The CG model considers a nucleotide by a single bead, then a contact map is built on the basis of the distances between the interaction sites in the native structure. Such is the approach used in the Self-Organized Polymer (SOP) model proposed by Hyeon and Thirumalai[25], designed to analyze the dynamics of RNA unfolding under constant force.

As in the CSU method, additional physicochemical details can be introduced by considering the interaction type between nucleotides. The main forces that stabilize RNA structures are due to stacking interactions and base pairing. The former is present when two nucleobases are close enough and lie on parallel planes exhibiting an overlap between their faces. On the other hand, base pairs originate from hydrogen bonds formed between the edges of the nitrogenous bases, yielding a relatively large number of possible geometrical arrangements between the four nucleobases that characterize RNA molecule: adenine (A), uracil (U), guanine (G), and cytosine (C). In particular, A-U or C-G base pairs and stacking interactions give rise to the well-known A-form double helix, a motif of extreme importance in RNA structure. Electrostatics can also be considered explicitly regardless of the proximity of the nucleotides in the native structure. For this purpose, a point charge is generally placed on the phosphorus atom on the backbone, which interacts with other charged particles through an implicit solvent approach.

Some Gō models have employed specific terms or functional forms for stacking and base pairs contacts using this information. The three interaction site model of Hyeon and Thirumalai[25,26] define a nucleotide by three point-particles representing the nucleobase (A, U, C, G), sugar ring (i.e. ribose: $C_5H_{10}O_5$), and phosphate group (**$PO_4^{3-}$**), which allows the introduction of directional interactions. The model has been parameterized with melting temperatures of small RNA fragments to study RNA folding thermodynamics under several ion concentrations and temperatures, and its interactions are also specific for contacts belonging to a double helix. Later versions of the model, however, are capable of introducing complementary base pairs between non-native contacts and stacking interactions between non-consecutive nucleotides[27]. This combination makes it possible to deal with a more complex free energy landscape, while introducing contacts which stabilize the native structure and taking care of describing properly the thermodynamics of the double helices which have an important contribution in the overall stability. In addition, the model of Hori and Takada[28], designed for the study of structural deformations of RNA and protein-RNA complexes, also uses a three-point representation of a nucleotide and a parametrization from MD simulations and distinguishes stacking from base-pairs in their Gō-like approach.

The large number of non-complementary base pairs and the possibility of forming hydrogen bonds between nucleobases and phosphate groups or sugar rings increases the complexity of the interaction network of RNA molecules. Despite this, several tools such as ClaRNA[28,29], FR3D[30], or x3dna-dssr[30,31] can be used to annotate structures and identify the most relevant interactions in the system of interest, which can help to build Go models able to capture the essentials for the phenomena to study under simulations.

## 3. The structure-based model: A Gō-like approach for protein-sugar complexes

In nature, proteins and polysaccharides can exist separately and also form complexes, for instance, the degradation of cellulose fibrils by fungi or bacteria involves processing of the biomaterial by enzymes (e.g. endo- and exo-glucanases) and thus a relevant biotechnological process that has been improved for the biofuel production[32,33]. At the molecular level, enzymes recognize the cellulose chain ends or broken chains and after attaching to it, the cleavage of the O-glycosidic bond is carried out, releasing several small oligomers that can be the source of energy for several microorganisms. Also, glycosylation of proteins by sugar moieties (i.e. N-glycan or O-glycan) can induce conformational changes via allosteric communication. Such an effect was reported in the conformational transition from closed to open state in the SARS-CoV-2 spike (S) glycoprotein[34]. The relevance of describing such events by molecular simulation can lead us to the development of novel therapeutics against pathogens such as viruses and bacteria. In this regard, the study of protein-sugar complexes remains an active field of research in the biomolecular community. The extension of the Gō-like approach for the study of protein-sugar complexes was carried out in ref.[35]. In this work, the Cα-based Gō-like model for proteins was coupled with an SB CG model for polysaccharides. Each sugar oligomer was formed by D-glucose units connected by the β(1→4) glycosidic bonds in cellohexaose whereas the α(1→4) glycosidic bonds in amylohexaose case. Then each sugar monomer in the CG description was represented by one CG bead centred on the position of the carbon atom. We considered the position of the C4 atom for comparison with respect to C1 position. Alternatively, we also derive parameters for C1 and the centre-of-mass of the monomer. The new set of CG values for bonded and non-bonded parameters were determined by AA-MD simulations using two statistical methods. One of these methods was the Boltzmann inversion (BI) method[36] while the other was denoted by the energy-based (EB) approach. The nonbonded parameters for the protein-sugar complex were mapped by the Lennard-Jones (12-6) potential according to the Gō-like approach. These two methods were employed to calculate the stiffness of sugar oligomers and protein secondary structures. Protein Man5B comprises 330 residues, with PDB ID 3W0K. This study shows the stiffness of α-helices, which on average is stiffer than β-strands. Also note that in proteins, the secondary structures are generally stiffer (based on elastic contacts) by a factor of 5 than in sugars. This CG model was validated for the case of hexaose-Man5B catalytic complex (see Fig. 2). The large fluctuations calculated by principal component analysis (PCA) of the active loop in Man5B were retained in the CG description. The main trend between AA-MD[37] and CG-MD simulations regarding the binding activity was also captured in the CG model.

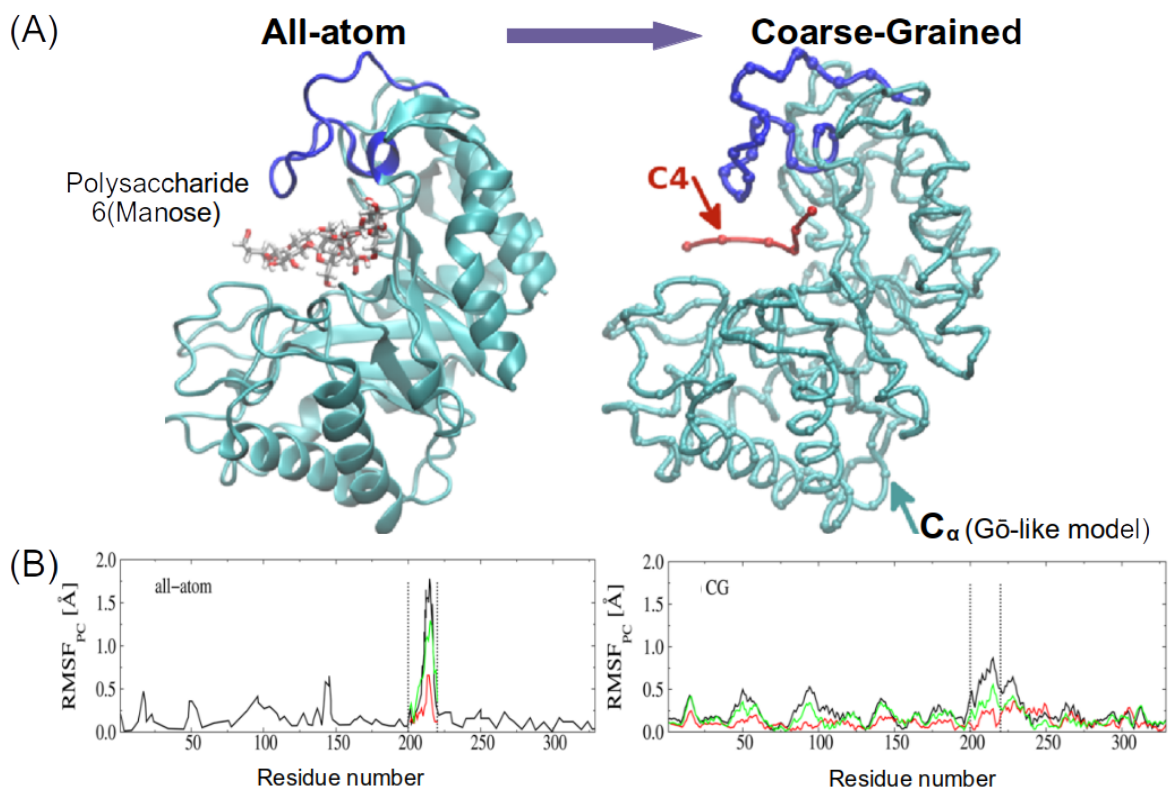

**Figure 2.** Panel A shows the all-atom MD and SB CG modelling of the sugar-Man5B complex (PDB: 3W0K). The blue protein segment comprising the residues (200-220) which is considered the active loop responsible for the cleavage of the O-glycosidic bond in polysaccharides. The CG model employs the $C_\alpha$ positions for protein and C4 atoms for the sugar hexamer. Panel B shows the fluctuations of the complex under AA and CG simulations. The RMSF shows fluctuations of the protein segment undocked (black line) and docked with manohexaose and cellohexaose in green and in red solid lines respectively. This figure was adapted with the permission from ref.[35]

These results highlighted the energetic differences between protein-sugar interactions and native interaction in proteins ($\epsilon_{PP}$ ~1.5 kcal/mol). It was reported the strength of sugar-protein energy value ($\epsilon_{SP}$) in the range of 3 to 6 kcal/mol. This SB model for protein-sugar complex is constructed under implicit solvent conditions and no detailed chemistry of residues is included, thus ligand recognition associated with long-range interaction or the effect of single point mutations that induce conformational changes cannot be captured by this simple model. Furthermore, atomistic backmapping is not doable under this representation because of the level of CG description based on $C^{\alpha}$ atoms. In the next section, we present an alternative approach that is based on the Gō-like model for proteins, and it circumvents several limitations of this SB model.

## 4. Overview of the GōMartini approach for protein complexes

The GōMartini approach was first introduced by Poma *et al*.[1] and it coupled the Martini 2 and with the SB model for protein (i.e. Gō-like type) developed by Cieplak's lab. The protocol for GōMartini approach is depicted in Fig. 3, the first step begins with an experimental 3D structure of a globular protein. From this structure, the OV+rCSU contact map is created from the server http://pomalab.ippt.pan.pl/GoContactMap/[38,39]. The next step involves the transformation from AA to CG representation using ./martinize2 script[40]. In the case of protein complexes, each of its chains must be isolated in individual PDB files. The GōMartini approach is applied to obtain a CG structure with the create_gomartini.py script. The strength of the Gō potential mapped by a Lennard-Jones(12-6) potential can be adjusted to match results from experiments or atomistic simulations. The GōMartini approach allows the sampling of large-scale conformational changes, a limitation in AA-MD simulation[1]. This holds particular significance in the investigation of proteins that undergo significant structural transitions, such as unfolding events under a mechanical force, inter-domain motions, and catalytic rearrangements, as well as a better description of protein complex stability [38,41–45].

The Martini force field allows for the simulation of biomacromolecules at a faster rate than the conventional AA representation. For proteins, an old SM model based on an elastic network (EN) model was used to preserve protein native structure, by adding harmonic bonds between $C^{\alpha}$ atoms. This has the drawback of not being able to explore the full conformational landscape of the protein, as well as overestimating the number of contacts between nearby residues[1]. The GōMartini approach involves replacing the contacts obtained by the EN model based on a simple cut-off distance by the native contacts built based on the OV and a chemistry-based rCSU CMs. Then, virtual sites are placed near the BB particles. Such implementation in Gromacs 2020 (or above) enhances the production stage. The only energy scale in this approach ($\epsilon_{G\bar{o}}$) can be modified iteratively. The default value is 9.414 kJ/mol which corresponds to the energy of the hydrogen bond in proteins[35]; however, alternative values have been employed to replicate experimental results, including 12 kJ/mol, and in certain instances, values of 100 kJ/mol over certain pairs of contacts were necessary[39]. Below is a brief review of successful GōMartini studies with Martini 2[47] and Martini 3[2] force fields.

*4.1 Martini 2*

The first applications of the GōMartini approach were carried out using the Martini 2 force field. It examined the folding of small α- and □-peptides, the conformational flexibility of a set of proteins, and the nanomechanics of a titin domain using different definitions of contact maps and different $\epsilon_{G\bar{o}}$ values. Folding simulations were done for an α-helix segment of the histidine-containing phosphocarrier protein and a □-strand of the G protein.

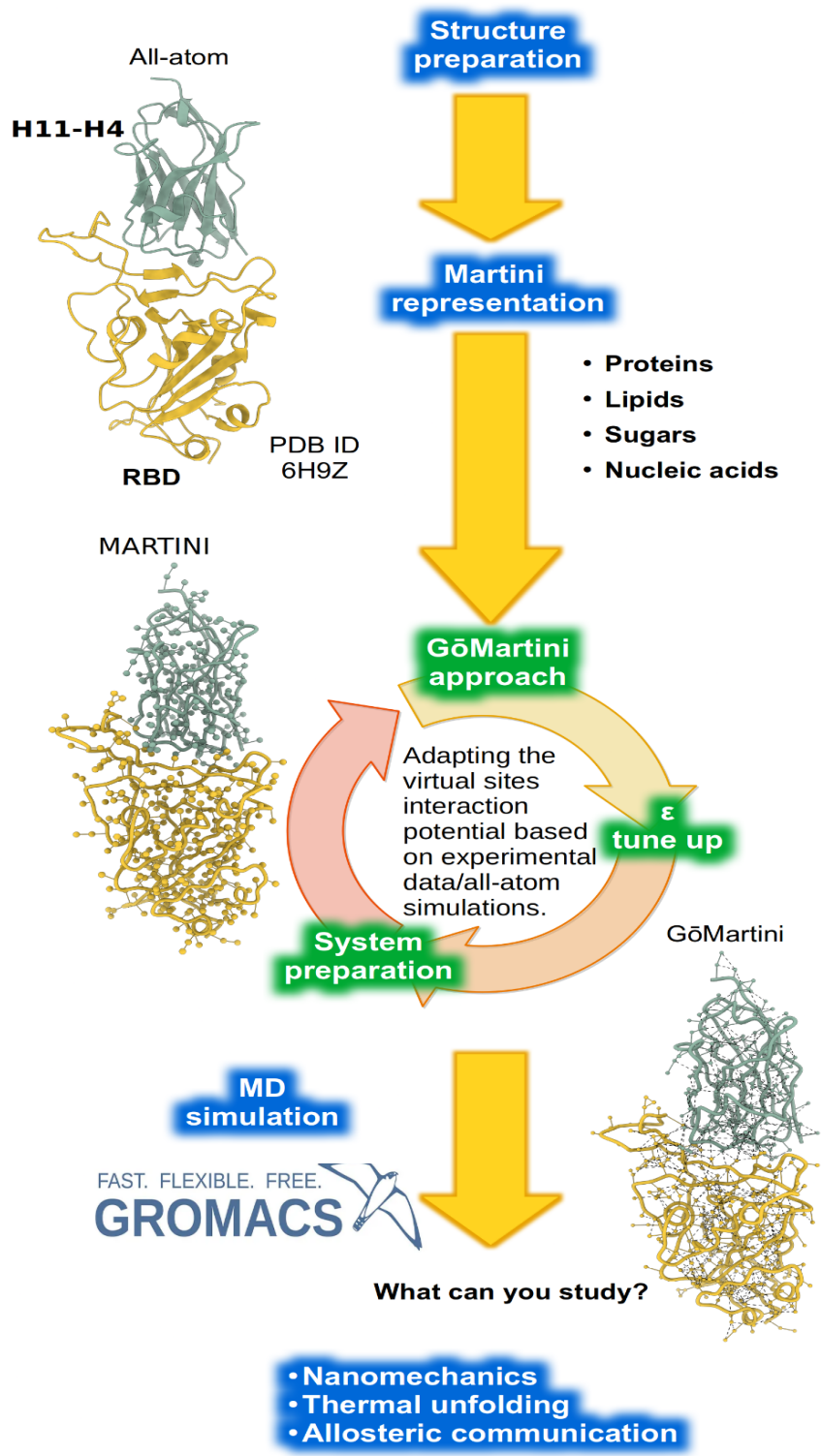

**Figure 3**. The GōMartini approach workflow for protein complexes. The study of oligomeric complexes involves building a Martini CG representation, creating a contact map, and introducing Gō bonds mapped as LJ potential through the virtual site implementation (denoted as dummy beads in GROMACS versions before 3.3). The resulting complex is solvated and neutralized at ambient conditions. MD simulations are performed using GROMACS MD package[46]. By iteratively modifying $\varepsilon_{Gō}$ value in the range of 9.4 and 12.0 kJ/mol, CG-Martini simulations are capable of reproducing experimental results.

*4.1 Martini 2*

The first applications of the GōMartini approach were carried out using the Martini 2 force field. It examined the folding of small α- and □-peptides, the conformational flexibility of a set of proteins, and the nanomechanics of a titin domain using different definitions of contact maps and different $\varepsilon_{Gō}$ values. Folding simulations were done for an α-helix segment of the histidine-containing phosphocarrier protein and a □-strand of the G protein. Native contacts were calculated from the PDB structures, and the coordinates for the unfolded conformers were obtained from a CG simulation at 500 K with implicit solvent. The results showed that both petides refolded in almost all simulations. The equilibrium dynamics of the type I cohesin domain , the domain of I27 from titin, and ubiquitin (PDB IDs 1AOH, 1TIT, and 1UBQ, respectively) were examined at both atomistic and CG resolutions. RMSD analyses revealed that the proteins were stable along GōMartini simulations with deviations smaller than 0.2 nm and that characteristic residue fluctuations were captured during the MD simulation, in agreement with the previous EN model (i.e. ELNEDIN approach[48]) and AA-MD simulations. A principal component analysis indicated that the GōMartini approach was able to capture the opening and closing motion of the Man5B glycoside hydrolase. The amplitudes were comparable to those observed in AA-MD simulations[37]. Finally, nanomechanical studies employing GōMartini approach on the domain of I27 from titin showed: i) the nanomechanics can be captured at slower pulling speeds than AA-MD simulations, and ii) unfolding forces were similar to experimental values when extrapolated to low loading rates.

In another study, the GōMartini approach was used to describe the membrane remodeling dynamics of the F-Bin/Amphiphysin/Rvs (F-BAR) protein Pacsin-1[38]. The conformation of Pacsin-1 was not maintained using the original definition of native contacts, namely OV+rCSU, due to the overstabilization of contacts between neighboring residues. The definition of native contacts was redefined. Consequently, it considered all i-th and (i+3)th residue pairs. Also, if the minimum distance between all heavy elements (i.e. N, C, and O atoms) is shorter than a distance threshold, a pair of residues, i-th and j>i+3rd, is considered to have a native contact. Throughout the simulations, lateral Pacsin-1:Pacsin-1 interactions were observed and correlated with the solved 3D structure. This optimization reproduced the structural and local fluctuations observed in AA-MD simulations.

The GōMartini approach has been used to investigate the association of lipids with various proteins[49–51]. In one of these studies, the conformational dynamics and the effect of oligomerization of npq2 Light-harvesting complex II (LHCII) on the association with lipids[52]. Another study examined the stability of LHCII in its monomeric and trimeric forms, the cofactor flexibility, and the impact of membrane composition[51]. Both studies demonstrate the usefulness of the GōMartini approach for describing the conformational flexibility of proteins, with results comparable to those obtained by experimental techniques or AA-MD simulations.

The stability and enzyme flexibility of proteomimetics in the presence of zinc metalloproteinase thermolysin was studied using GōMartini. The simulation results were consistent with experimental observations[50]. In another study, the structural stability of the enzyme degradation PET-plastic (i.e. PETase) in a complex with copolymers at high temperatures was examined. The results obtained from GōMartini simulation were in agreement with the temperature-dependent conformation observed in AA-MD simulations[53].

One of the most notable applications of the GōMartini approach is in the nanomechanics of proteins that requires the use of steered molecular dynamics (SMD) simulations. The level of CG in this approach has the advantage of reaching experimental time and length scales, while maintaining a detailed description of the system at the molecular level[26]. In this particular aspect, several studies have dealt with the nanomechanics of Aβ$_{40}$, Aβ$_{42}$, α-synuclein, and other self-assembly peptides[42,44,45]. These investigations have shed light on the stability of biological fibrils[54] and their significance in the progression of neurodegenerative diseases, as well as on the mechanical properties that can be used to develop new materials with industrial uses[42,44,45]. In another study, the unbinding pathways of the complex anticalin:CTLA-4 and its nanomechanics under various pulling geometries, which led to diverse force-distance profiles, were investigated using AFM-single-molecule force spectroscopy (SMFS) experiments and GōMartini simulations. As a result, this approach explained the observed experimental patterns of mechanical stability that were attributed to pulling geometries and to the loss of native contacts between secondary motifs[39,43].

*4.2 Martini 3*

The recent version of Martini 3 for proteins[2] and other polysaccharides[55–57] has improved the ability of GōMartini for the study of large conformational transitions in protein under several environmental conditions. The protein copper, zinc superoxide dismutase was the first use of GōMartini and Martini 3 for the study of conformational events[41]. The authors captured the allosteric effect of the G93A mutation on the electrostatic loop (EL). Note that a larger flexibility of EL causes the opening of this loop, which further destabilizes the zinc binding site of this enzyme via an increase in the hydration levels. In accordance with hybrid quantum-mechanical/molecular-mechanical MD simulations, the opening of the EL was reproduced using simulations, as well as its conformational flexibility. This study paved the way for the utilization of the GōMartini methodology in the comprehensive examination of mutations and their allosteric effects on the structure and function of proteins. The implementation of CG-MD simulations employing the GōMartini strategy resulted in the identification of a second phosphatidylinositol 4,5-bisphosphate binding site on the C-terminal domain of the tubby protein. The validation of this new binding site was carried out by mutating charged residues to alanine, both *in silico* and in living cells. It was shown that the affinity for phosphoinositide was reduced in both experiments[58].

In a study of the accessory factors UbiJ and UbiK, the GōMartini approach was used to improve the sampling process. Also, the absorption of a trimeric protein in the membrane was studied. A contact profile along an AA-MD simulation between the protein and the membrane was necessary to tune the CG model, which improved the accuracy of the interactions[59]. Small bifunctional molecules capable of modulating protein-membrane interactions were studied by GōMartini[60]. The CorA transport system asymmetric gating mechanism was investigated using the same method. For this purpose, both AA-MD and CG-MD simulations with different conformations of the protein chain were performed. The highly dynamic conformational changes observed in the set of simulations were consistent with recent structural studies. Based on previously reported information and results from the CG simulations, the authors proposed a patent on the novel asymmetric gating model for this protein system[61].

The giant mechanical stability of the adhesion bone sialoprotein-binding protein (Bbp) of *Staphylococcus aureus* and its role in biofilm formation have been recently investigated using AA-MD and GōMartini SMD simulations. Single-molecule force spectroscopy[43] has given evidence to such a high-degree of mechanostability in Bbp. Additional experiments on the Bbp-fibrinogen-α complex revealed that this is one of the most mechanostable protein complexes studied so far. These results agreed with experimental SMFS data[43].

The GōMartini approach has proven to be useful for the study of diverse protein systems, revealing details about their nanomechanics, allosteric effects, and a deeper appreciation of their conformational flexibility. It is possible to extend this approach to the study of protein complexes with diverse oligomeric states by using the workflow depicted in Fig. 3. GōMartini can compensate for protein-protein interactions that cannot be fully captured by the Martini 3 force field, enhancing our tools for the analysis of these complexes.

## 5. Perspective and conclusions

An interesting alternative for the study of biomacromolecular events at the nanometric scale and with a temporal resolution closer to experimental studies is presented by the GōMartini approach. The scope of its application will be widened with its future expansion to encompass other types of molecules, such as carbohydrates, lipids and nucleic acids. The GōMartini has become the gold-standard in Martini 3, as it offers flexibility by combining physical and chemical information in the construction of the contact map in protein. This is a particular edge that renders the information stored in the Gō interaction crucial for the understanding of the mechanism of protein-protein dissociation as well as during the nanomechanical deformation, as one can track directly the rupture of Gō bonds as it will be in a continuum system. Martini 2 used to overestimate the protein-protein interaction and in the Martini 3, the protein interface requires the contribution of additional Gō bonds that can be obtained by the OV+rCSU CM. We anticipate that a combination between statistical potentials and machine learning approaches can assist the contact map determination in protein complexes.

## 6. Acknowledgements

A.B.P acknowledges Marek Cieplak for inspiring the research on structure-based models in protein-sugar complexes. R.A.M.S acknowledges Marek Cieplak in sharing the source code of the OV+rCSU contact map in Fortran. A.B.P. acknowledges financial support from the National Science Center, Poland, under grant No. 2022/45/B/NZ1/02519. M.Ch. acknowledges Marek Cieplak for encouraging investigation on structures containing cavities, knots, and various unstructured systems, explored using coarse-grained models. M.Ch. acknowledges financial support from the National Science Centre (NCN), Poland, under grant No. 2018/31/B/NZ1/00047 and the European H2020 FETOPEN-RIA-2019-01 grant PathoGelTrap No. 899616. A.B.P and M.Ch acknowledge support of the computer resources by the PL-GRID infrastructure.

TOC image

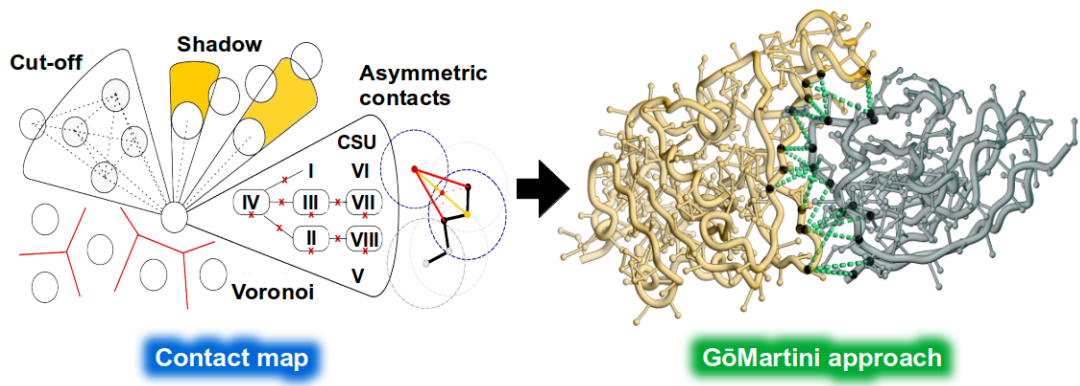